# Log-periodic quantum magneto-oscillations and discrete scale invariance in topological material HfTe$_5$

Huichao Wang,[1,2] Yanzhao Liu,[1] Yongjie Liu,[3] Chuanying Xi,[4] Junfeng Wang,[3] Jun Liu,[5] Yong Wang,[5] Liang Li,[3] Shu Ping Lau,[2] Mingliang Tian,[4] Jiaqiang Yan,[6] David Mandrus,[6,7] Ji-Yan Dai,[2,*] Haiwen Liu,[8,‡] X. C. Xie,[1,9,10] and Jian Wang[1,9,10,†]

[1]*International Center for Quantum Materials, School of Physics, Peking University, Beijing 100871, China*
[2]*Department of Applied Physics, The Hong Kong Polytechnic University, Kowloon, Hong Kong, China*
[3]*Wuhan National High Magnetic Field Center, Huazhong University of Science and Technology, Wuhan 430074, China*
[4]*High Magnetic Field Laboratory, Chinese Academy of Sciences, Hefei 230031, Anhui, China*
[5]*Center of Electron Microscopy, State Key Laboratory of Silicon Materials, School of Materials Science and Engineering, Zhejiang University, Hangzhou, 310027, China*
[6]*Materials Science and Technology Division, Oak Ridge National Laboratory, Oak Ridge, Tennessee 37831, USA*
[7]*Department of Materials Science and Engineering, University of Tennessee, Knoxville, Tennessee 37996, USA*
[8]*Center for Advanced Quantum Studies, Department of Physics, Beijing Normal University, Beijing, 100875, China*
[9]*Collaborative Innovation Center of Quantum Matter, Beijing 100871, China*
[10]*CAS Center for Excellence in Topological Quantum Computation, University of Chinese Academy of Sciences, Beijing 100190, China*

Abstract

Discrete scale invariance (DSI) is a phenomenon featuring intriguing log-periodicity which can be rarely observed in quantum systems. Here we report the log-periodic quantum oscillations in the magnetoresistance (MR) and the Hall traces of HfTe$_5$ crystals, which reveals the appearance of DSI. The oscillations show the same log$B$-periodicity in the behavior of MR and Hall, indicating an overall effect of the DSI on the transport properties. Moreover, the DSI feature in the Hall resistance signals its close relation to the carriers. Combined with theoretical simulations, we further clarify the origin of the log-periodic oscillations and the DSI in the topological materials. Our work evidences the universality of the DSI in the Dirac materials and paves way for the full understanding of the novel phenomenon.

Discrete scale invariance (DSI) is a partial breaking of continuous scale invariance where observables of the system obey the scale invariance only for a geometrical set of choices written as the form of $\lambda^n$ with $\lambda$ being the scaling ratio [1]. With the violation of the classical continuous scale symmetry, the DSI represents a scale anomaly, and the characteristic signature of DSI, the intriguing log-periodicity, exists in rupture, growth processes, turbulence, finance, and so on. The appearance of log-periodic structures indicates the characteristic length scales in a system, which is extremely interesting when it is fundamentally related to the underlying physical mechanism [1].

The scale anomaly DSI is of high general interest while it can be rarely observed in quantum systems experimentally [2]. For a long time, the DSI has only been confirmed in cold atom systems and generated tremendous interest [3-10]. Nowadays, the DSI behavior in Dirac materials has also attracted attention in several subfields of physics [11-16]. Especially, the magneto-transport measurements on topological material ZrTe$_5$ reveal a new type of magnetoresistance (MR) oscillations with peculiar log-periodicity and thus manifest the appearance of DSI in a solid-state system [14]. Such peculiar DSI feature is considered to be universal in Dirac materials with Coulomb attraction [14], which may be closely related to the quasi-bound states formed by massless Dirac fermions and the long pursued atomic collapse phenomenon [14-15,17]. Thus, it is desirable to explore the log-periodic quantum oscillations and the DSI in other physical observables and in other Dirac systems to justify the universality and provide insights into the underlying mechanisms.

$HfTe_5$ is a sister compound of $ZrTe_5$ [18-21]. In recent years, the theoretical prediction of topologically nontrivial nature of these two materials has triggered a new wave of research boom [22]. Lately, the electrical transport measurements have revealed that $HfTe_5$ shows ultrahigh mobility [23], indicating great promise in electronic applications. In addition, the negative MR under parallel magnetic and electrical fields is observed in $HfTe_5$ [23,24], which could be attributed to the chiral anomaly [25,26] or the nontrivial Berry curvature [27,28]. The pressure induced superconductivity in $HfTe_5$ also illustrates interesting property of the system [29,30]. According to the first-principles calculations, the pentatelluride compound is extremely sensitive to the cell volume and can manifest as a Dirac semimetal or a topological insulator under variant crystal volumes [31], which depends on the sample quality [32]. Thus, the pentatelluride compound has become a promising topological material platform.

In this work, we reveal the universality of the peculiar log-periodic quantum oscillations and DSI phenomenon in Dirac materials by the magneto-transport results of $HfTe_5$ crystals. The oscillations with log$B$-periodicity are demonstrated in the MR behavior, independent on the minor differences of the sample quality. The log$B$-periodicity is also discovered in the Hall traces of the $HfTe_5$ crystals. This indicates an overall effect of the DSI on the transport properties. Based on the experimental results, we elucidate the origin of the DSI, and elaborate on its relation to the log-periodic oscillations in both the longitudinal MR and the Hall resistance. We propose a theoretical model to quantitatively explain these features, analyze the influence of a small band gap on the DSI feature and clarify its relevance to various topological materials. This work provides new insights towards further understanding of the log-periodic quantum oscillations and the DSI in solid-state systems.

Single crystals $HfTe_5$ in our work were grown via a self-Te-flux method as in the previous report [23]. The crystals were chemically and structurally analyzed by powder X-ray diffraction, scanning electron microscopy with energy dispersive X-ray spectroscopy, and transmission electron microscopy. The atomically high-resolution transmission electron microscopy image of one typical sample is shown in the inset of Fig. 1(a), which manifests a high-quality nature. Electrical transport measurements in this work were conducted in three systems, a 16 T- PPMS (Physical Property Measurement System) from Quantum Design, a pulsed high magnetic field facility (58 T) at Wuhan National High Magnetic Field Center and a static magnetic field up to 25 T in the High Magnetic Field Laboratory in Hefei. Results from different measurement systems and different samples are reproducible and consistent with each other. Standard four/six-electrode-method was used for the MR/Hall measurements with the excitation current flowing along the crystallographic $a$ axis of $HfTe_5$.

The resistivity-temperature ($\rho T$) characteristic of $HfTe_5$ crystals down to 2 K is shown in Fig. 1(a). With decreasing temperatures, the samples show firstly the metallic behavior above approximate 200 K and then a semiconducting-like upturn. As the temperature is further decreased, a sample dependent resistivity peak is observed at temperatures $T_p$ varying from 20 K to 40 K. At even lower temperatures, the semiconducting-like upturn recovers in most samples. It is noted here that the resistivity peak in the crystals cannot be attributed to Lifshitz transition since the Hall remains positive up to room temperature with no sign change [23].

Figure 1(b) shows the MR behavior at 2 K of different samples from the same batch when the magnetic field is perpendicular to the layer orientation (B//$b$ axis). The MR follows a sharp cusp at around zero magnetic field and becomes saturated at high magnetic fields. The MR ($R(H)/R(0)$) values show sample dependence with a range of 1500 %-5500 % at 15 T. According to the nonlinear Hall data of HfTe$_5$ crystals, we would attribute the various $\rho T$ behavior and MR effect to the competition of a semi-metallic Dirac band and a semiconducting band in the material [33]. For the sample 3 (s3), the Fermi level is very close to the Dirac point and thus the Fermi surface of the Dirac pocket is tiny. The dominated semiconducting band induces the upturn resistance at low temperatures and the MR effect is small. To the contrary, the Dirac band dominates in s4, which gives rise to the metallic $\rho T$ behavior at low temperatures down to 2 K and the larger MR effect in s4. For the other two samples s1 and s2, they exhibit the intermediate properties as shown in Figs. 1(a) and 1(b). Since the transition-metal pentatelluride system is extremely sensitive to the cell volume, the slight diversity of these transport properties could be attributed to the minor quality differences in the samples. Similar two-band model was ever proposed to interpret the mysterious peak in ZrTe$_5$ considering either Te deficiency or Iodine contamination [34,35]. The model might give a unified explanation for the various observations of the Dirac or semiconducting property in ZrTe$_5$.

By performing the second derivative for the MR results in Fig. 1(b), oscillations can be distinguished from the large MR background. The characteristic magnetic fields $B_n$ of oscillating peaks (marked with index $n$) and dips ($n$-0.5) in the oscillations is approximately consistent for different HfTe$_5$ samples (Fig. 1(c)). By plotting log$B_n$ vs. $n$ in Fig. 1(d), the index dependence for different samples can all be reproduced by a linear fitting, which reveals peaks and dips appear periodically as a function of log$B$ in HfTe$_5$. We identify that these specific magnetic fields satisfy the law of $B_n=\lambda B_{n+1}$, where $\lambda$ is a characteristic scale factor for the material. From the index plot in Fig. 1(d), the dominant scale factor $\lambda$ is shown to be 2.5 or 3.0.

The magneto-transport measurements at high magnetic fields up to 58 T further confirm the log$B$-periodic MR oscillations and DSI in HfTe$_5$. For clarity, data curves in Figs. 2(a) and 2(b) are shifted. The pink curve for s5 in Fig. 2(a) is measured at a static magnetic field. The MR oscillations observed at lower magnetic fields can be well reproduced by the pulsed magnetic field measurements on s5 and more oscillations are observed at higher magnetic fields (orange and red), as guided by the dashed lines. Besides, the oscillations are also observed in other samples s6 and s7, where the resistance peaks and dips in s5 can be replicated. The oscillations can be extracted by subtracting a smooth background from the raw data in Fig. 2(a) and the results are shown in Fig. 2(b). The index plot for the oscillations are shown in Fig. 2(c), which confirms the log$B$-periodicity of the MR oscillations with more experimental points at ultrahigh magnetic fields (green dots). By performing the Fast Fourier Transform (FFT) of the log-periodic oscillations in Fig. 2(b), a sharp FFT frequency peak is observed in various samples (Fig. 2(d)), which is consistent with the linear fitting results shown in Fig. 2(c). It is worth noting that the factor $\lambda$ has a broadening with a width in experiments. Based on an error bar determined by the full width at half maximum (FWHF) of the FFT frequency peak and combining with the results of different samples, we obtain a factor range of about [2.5, 5.9] in the HfTe$_5$ crystals.

The temperature dependence of the log$B$-periodic oscillations in HfTe$_5$ is shown in Fig. 3. Figures 3(a)-3(c) are results for s5 and Figs. 3(d)-3(f) are results for another sample s7. By subtracting background from the raw MR data in Figs. 3(a) and 3(d), we obtain the oscillating resistance shown in Fig. 3(b) and Fig. 3(e) respectively and then perform FFT of the oscillations. The FFT amplitudes in Figs. 3(c) and 3(f) are normalized divided by the peak amplitude at the base temperature. Based on the theoretical formula of $A=A_0(1-\exp(-\Delta E/k_BT))$, the fitting of the FFT amplitude at varied temperatures gives a characteristic binding energy

Δ$E$ of 7.0 meV for s5 and 7.5 meV for s7. The binding energies determine the disappearance temperatures $T_d$ of the oscillations, with the fitting value of $T_d$ equaling to 81 K and 87 K, respectively. The characteristic temperatures for both samples are consistent with our experimental observations.

We finally investigated the influence of the DSI on the Hall traces of HfTe$_5$. Since the noise at pulsed magnetic field is large, it is hard to obtain good Hall signals. Thus, the Hall data is measured in two different measurement systems with static magnetic fields, a 16 T-PPMS and a facility in the High Magnetic Field Laboratory in Hefei. In Fig. 4(a), the Hall data of HfTe$_5$ clearly shows a non-linear dependence on the magnetic field, which is consistent with the two-band model. Distinct and consistent oscillations are observed in the Hall resistance of different samples. The second derivate results are shown in Fig. 4(b). Similar to the property of oscillations on the MR, the characteristic log$B$-periodicity of the Hall resistance oscillations is confirmed by the linear index dependence (Fig. 4(c)). We also perform FFT of the oscillating Hall resistance in s8 and a range of [2.4, 5.9] can be obtained (Fig. 4(d)), which is in accordance with the result of the MR oscillations. Thus, the log-periodic quantum magneto-oscillations in both the longitudinal MR and the Hall traces indicate the underlying DSI property of the topological material HfTe$_5$ and its evolution under the magnetic field.

As discussed previously [14], the log-periodic quantum magneto-oscillations cannot be attributed to the conventional quantum oscillations, such as the Shubnikov–de Haas oscillations even with the consideration of the Zeeman-effect. In addition, the peculiar phenomenon shows different feature compared with the field-induced Fermi surface deformation or reconstruction scenario, such as the density-wave transition. It is suggested that log-periodic oscillations are closely related to the Weyl particles from the hole band with long-range Coulomb attraction when the carrier density is so dilute, and the long-range Coulomb attraction is generated by the charge impurity or the opposite type of carriers [14].

The Weyl equation with Coulomb attraction $V(\vec{R}) = \frac{-Ze^2}{4\pi\varepsilon_0 R}$ obeys the scale invariance property [14,15]. Here $Ze$ is the central charge, and the fine structure constant $\alpha = \frac{e^2}{4\pi\varepsilon_0 \hbar v_F}$ is of order one due to the small Fermi velocity $v_F$ in Dirac materials. Thus, the so-called supercritical condition ($Z \cdot \alpha$ surpassing the angular momenta $\kappa$) can be matched in the system, and further gives rise to quasi-bound states solution in the system. In the following, we only consider the lowest angular momentum channel with $\kappa = 1$. The combination of scale invariance and quasi-bound state solution results in the DSI property, and the radius of the quasi-bound states satisfy the relation $\frac{R_{n+1}}{R_n} = e^{\pi/s_0}$ with $s_0 = \sqrt{(Z \cdot \alpha)^2 - 1}$ [14].

When a magnetic field is applied, the magnetic length $l_B = \sqrt{\hbar c/eB}$ is included, and the binding energy spectrum of the quasi-bound states evolve with the magnetic field. Using the Wentzel-Kramers-Brillouin method, our numerical simulation shows that the energy of the $n$-th quasi-bound states approaches the Fermi energy at the magnetic field $B_n$, which obeys the approximate DSI property [14]. Aside from the quasi-bound states near the Coulomb center, large number of mobile carriers also exist in the lowest Landau level under the magnetic field in the ultra-quantum limit. Thus, the resonant scattering between the mobile carriers and the quasi-bound states around the Fermi level determine the transport properties of the material, e.g. the longitudinal MR and the Hall traces [36].

The oscillation term in $\rho_{xx}$ and $\rho_{yx}$ can be obtained by the T-matrix approximation, and both satisfy the log-periodic property [36]. The correction to $\rho_{xx}$ is of the order $n_C/n_S$, while that to $\rho_{yx}$ is of the order $n_C/N$, here $n_C$, $n_S$ and $N$ denote the density of the quasi-bound states, the density of short-range impurity and the total carrier density, respectively. Thus, the log-periodic magneto-oscillations only occupy small percentage of total resistance due to small ratio of $n_C/n_S$ and $n_C/N$. In HfTe$_5$, the oscillating resistance is about 0.4-0.9% of the longitudinal MR for the n=2 peak as shown in Fig. 3, and the oscillations in the Hall resistance is more apparent (about 2%) as shown in Fig. 4. When comparing the results of HfTe$_5$ with those of ZrTe$_5$ [14], we find two differences. Firstly, the oscillating amplitude of the log-periodic quantum oscillations in $\rho_{xx}$ of HfTe$_5$ is relatively smaller. We attribute the phenomena to a large $n_S$ and relatively small $n_C/n_S$, thus the correction to $\rho_{xx}$ is weak. Meanwhile, due to the relatively large impurity scattering from these short-range impurities, the broadening effect smears the log-periodic oscillations at small magnetic fields, ultimately leading to fewer observable oscillating cycles. Secondly, the log-periodicity in $\rho_{yx}$ of HfTe$_5$ is relatively more remarkable. This feature may be due to the relatively large value for the density of charge impurity and thus relatively large value for the density of quasi-bound states $n_C$ and the ratio $n_C/N$, which gives rise to considerable correction for Hall trace $\rho_{yx}$ in HfTe$_5$.

Besides the ideal consideration shown above, in real systems some other issues can influence the DSI property, which needs to be clarified. Firstly, both a small band gap $\Delta$ and finite screening length $\lambda_S$ impose constraint on the DSI feature by introducing an effective low energy cut-off. But the quasi-bound states with binding energy much larger than the band gap $\Delta$ and radius much smaller than $\lambda_S$ cannot be influenced. Direct estimation gives the largest value of radius $R_C$ and corresponding magnetic length $l_{B,C}$ satisfying $R_C \approx l_{B,C} \approx$ min ($\hbar v_F/\Delta$, $\lambda_S$). If one set $\Delta = 10$ meV and neglect the screening effect, $R_C \approx 33$ nm and the corresponding magnetic field $B_C \approx 0.6$ T. Secondly, in the system with ultralow carrier density, both the charge impurity and the carrier from the electron band can generate the Coulomb attraction [16]. Finally, although this background subtraction procedure is generally used and accepted for the analysis of magneto-oscillations, the previous analysis of DSI usually utilize the log-log plot. Here, out of the log-periodic quantum oscillations, other scattering mechanisms also contribute to the magneto-transport with a non-oscillating background. The theoretical formulas reveal similar results, in which only the oscillations obey the DSI feature [36]. Thus, the commonly used procedure of background subtraction for magneto-oscillations is utilized to demonstrate the DSI feature in our results [37]. The DSI feature has been clearly revealed in the MR and the Hall results of the HfTe$_5$ crystals by the distinct log-periodicity and the theoretical comparison [36].

In summary, we report the intriguing log-periodic quantum magneto-oscillations in both the Hall and the MR results in the topological material HfTe$_5$, which indicate the underlying DSI feature in the system. The observation of the log-periodicity in both physical observables $\rho_{xx}$ and $\rho_{yx}$ reveals that the DSI shows an overall effect on the transport properties for a Dirac system with the long-range Coulomb attraction. The origin of the DSI and its relation to the log-periodic magneto-oscillations are theoretically elucidated. This work paves the way for further research on the log-periodic oscillations and the DSI in quantum systems.

We acknowledge Zhibo Dang for discussions on the data. This work was financially supported by the National Basic Research Program of China (Grants No. 2018YFA0305604, Grants No. 2017YFA0303300, No. 2015CB921102), the National Natural Science Foundation of China (Grants No.11774008, No. 11674028, No. 11534001, No. 11504008), the Strategic Priority Research Program of Chinese Academy of Sciences (Grant No. XDB28000000), the Fundamental Research Funds for the Central Universities, Hong Kong GRF grant (153094/16P), the Hong Kong Polytechnic University strategic plan (No: 1-ZE25 and 1-ZVCG) and the Postdoctoral Fellowships Scheme (No. 1-YW0T). Work at ORNL was supported by the US Department of Energy, Office of Science, Basic Energy Sciences, Division of Materials Sciences and Engineering.

H.W. and Y. L. contributed equally to this work.

*jiyan.dai@polyu.edu.hk;
‡haiwen.liu@bnu.edu.cn;
†jianwangphysics@pku.edu.cn
[1] D. Sornette, *Discrete-scale invariance and complex dimensions*, Phys. Rep. **297**, 239-270 (1998).

[2] L. D. Landau and E. M. Lifshitz, *Quantum Mechanics: Non-relativistic Theory*, 3rd ed. (Perganon Press, Oxford, U.K., 1977).

[3] V. Efimov, *Energy levels arising from resonant two-body forces in a three-body system*, Phys. Lett. B **33**, 563 (1970).

[4] E. Braaten and H. W. Hammer, *Universality in few-body systems with large scattering length*, Phys. Rep. **428**, 259 (2006).

[5] P. Naidon and S. Endo, *Efimov physics: a review*, Rep. Prog. Phys. **80**, 056001 (2017).

[6] T. Kraemer *et al.*, *Evidence for Efimov quantum states in an ultracold gas of caesium atoms*, Nature **440**, 315-318 (2006).

[7] B. Huang, L. A. Sidorenkov, R. Grimm, and J. M. Hutson, *Observation of the Second Triatomic Resonance in Efimov's Scenario*, Phys. Rev. Lett. **112**, 190401 (2014).

[8] R. Pires *et al., Observation of Efimov Resonances in a Mixture with Extreme Mass Imbalance*, Phys. Rev. Lett. **112**, 250404 (2014).

[9] S.-K. Tung, K. Jiménez-García, J. Johansen, C. V. Parker, and C. Chin, *Geometric Scaling of Efimov States in a $^6$Li-$^{133}$Cs Mixture*, Phys. Rev. Lett. **113**, 240402 (2014).

[10] M. Kunitski *et al.*, *Observation of the Efimov state of the helium trimer*, Science **348**, 551-555 (2015).

[11] W. Greiner, B. Muller, and J. Rafelski, *Quantum Electrodynamics of Strong Fields* (Springer-Verlag, Berlin, 1985).

[12] A. V. Shytov, M. I. Katsnelson, and L. S. Levitov, *Atomic Collapse and Quasi–Rydberg States in Graphene*, Phys. Rev. Lett. **99**, 246802 (2007).

[13] Y. Nishida, *Vacuum polarization of graphene with a supercritical Coulomb impurity: Low-energy universality and discrete scale invariance*, Phys. Rev. B **90**, 165414 (2014).

[14] H. Wang *et al.*, *Discrete scale invariance and fermionic Efimov states in ultra-quantum ZrTe$_5$*, arXiv: 1704.00995.

[15] O. Ovdat, Jinhai Mao, Yuhang Jiang, E. Y. Andrei and E. Akkermans, *Observing a scale anomaly and a universal quantum phase transition in graphene*, Nat. Commu. **8**, 507 (2017).

[16] P. Zhang and H. Zhai, *Efimov Effect in the Dirac Semi-metals*, Front. Phys. **13**(5), 137204 (2018).

[17] Y. Wang *et al.*, *Observing atomic collapse resonances in artificial nuclei on graphene*. Science **340**, 734-737 (2013).

[18] S. Furuseth, L. Brattas, and A. Kjekshus, *Crystal Structure of HfTe$_5$*, Acta. Chem. Scand. **27**, 2367 (1973).

[19] M. Izumi, K. Uchinokura, and E. Matsuura, *Anomalous electrical resistivity in HfTe$_5$*, Solid State Commun. **37**, 641 (1981).

[20] G. N. Kamm, D. J. Gillespie, A. C. Ehrlich, D. L. Peebles, and F. Levy, *Fermi surface, effective masses, and energy bands of HfTe$_5$ as derived from the Shubnikov–de Haas effect*, Phys. Rev. B **35**, 1223 (1987).

[21] T. M. Tritt, N. D. Lowhorn, R. T. Littleton, A. Pope, C. R. Feger, and J. W. Kolis, *Large enhancement of the resistive anomaly in the pentatelluride materials HfTe$_5$ and ZrTe$_5$ with applied magnetic field*, Phys. Rev. B **60**, 7816 (1999).

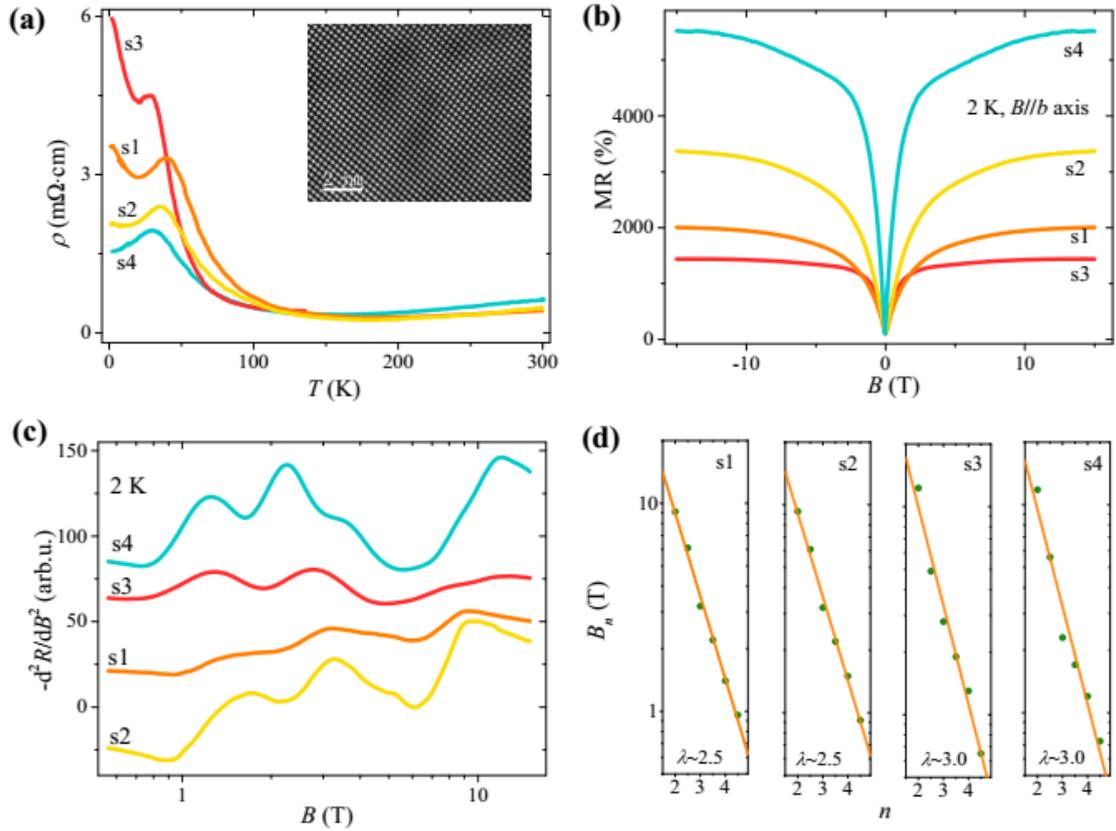

FIG. 1 (color online). Resistivity-temperature characteristic and MR behavior of HfTe$_5$ single crystals. (a) Temperature dependence of the resistivity. Inset: the atomically high-resolution transmission electron microscopy image of HfTe$_5$ manifesting a high-quality nature. (b) MR at 2 K for HfTe$_5$ crystals in a perpendicular magnetic field. (c) The second derivative results of the MR behavior at 2 K (Fig. 1(b)). Data curves are shifted for clarity. (d) Linear dependence of log$B_n$ on the index $n$ shows log-periodicity of the MR oscillations. $B_n$ is the characteristic magnetic field for a peak or dip in the oscillations.

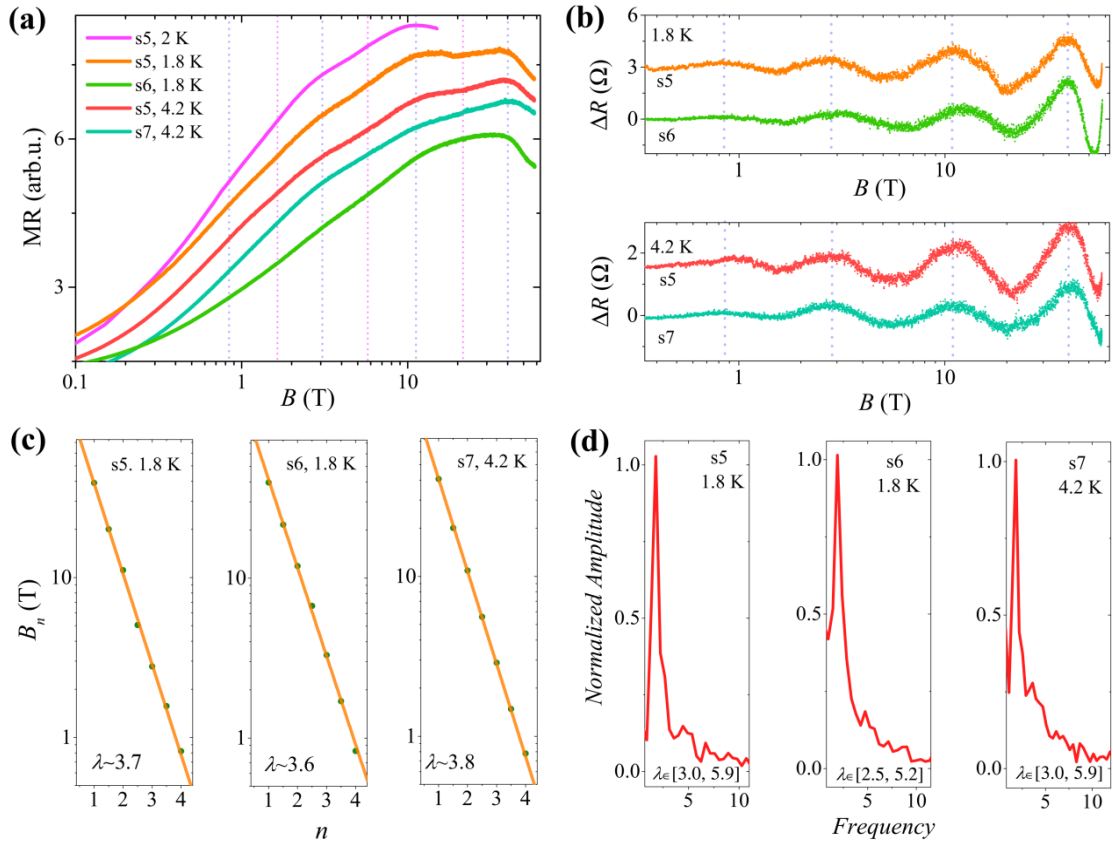

FIG. 2 (color online). Log-periodic MR oscillations in HfTe$_5$. (a) MR of HfTe$_5$ vs. log$B$. The MR oscillations measured in PPMS (pink) is consistent with the results (orange and red) in the pulsed high magnetic field. MR oscillations are reproduced in different samples (s5, s6, s7). Dashed lines serve as guides to the eye. (b) Extracted MR oscillations from the raw data in (a) after subtracting a background. Data curves in (a) and (b) are shifted for clarity. (c) Log$B$-periodicity of the MR oscillations in HfTe$_5$. (d) FFT results of the MR oscillations in (b). Combining the results of different samples, the scale factor $\lambda$ shows a range of about [2.5, 5.9] which is determined by the FWHM of the FFT frequency peak.

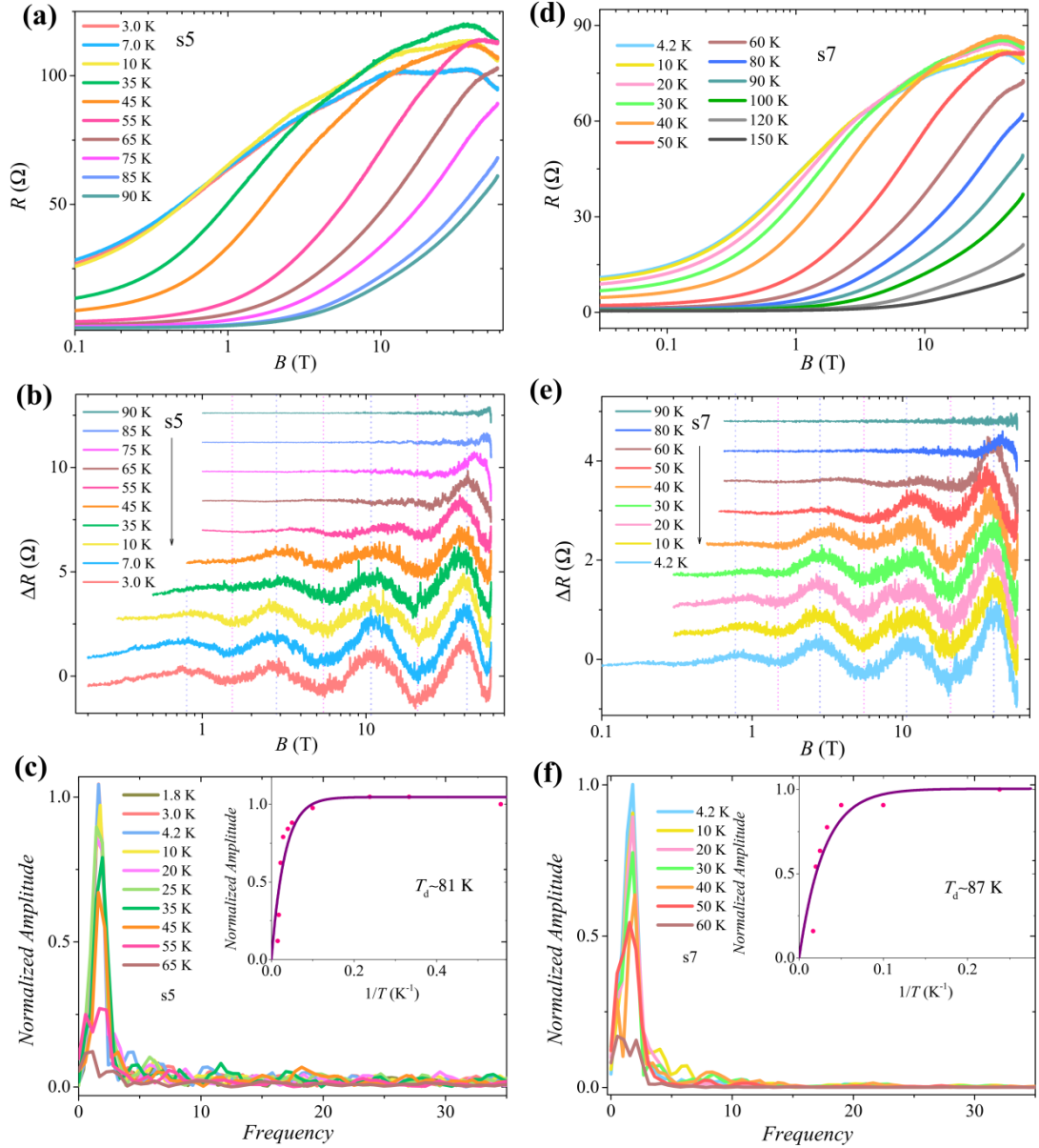

FIG. 3 (color online). Temperature dependence of the log$B$-periodic oscillations. (a) MR of HfTe$_5$ (s5) at selected temperatures. (b) Log$B$-periodic oscillations in s5 at selected temperatures. (c) FFT results for the MR oscillations in (b). Inset: theoretical fit on the normalized FFT amplitude at varying temperatures. The fitted value of the disappearance temperature $T_d$ of the oscillations is consistent with the experimental observations. (d)-(f) are results for another sample s7. The fitting parameter $T_d$ is also consistent with experiments. The characteristic temperature $T_d$ for HfTe$_5$ crystal is about 80-90 K. Data curves in (b) and (e) are shifted for clarity.

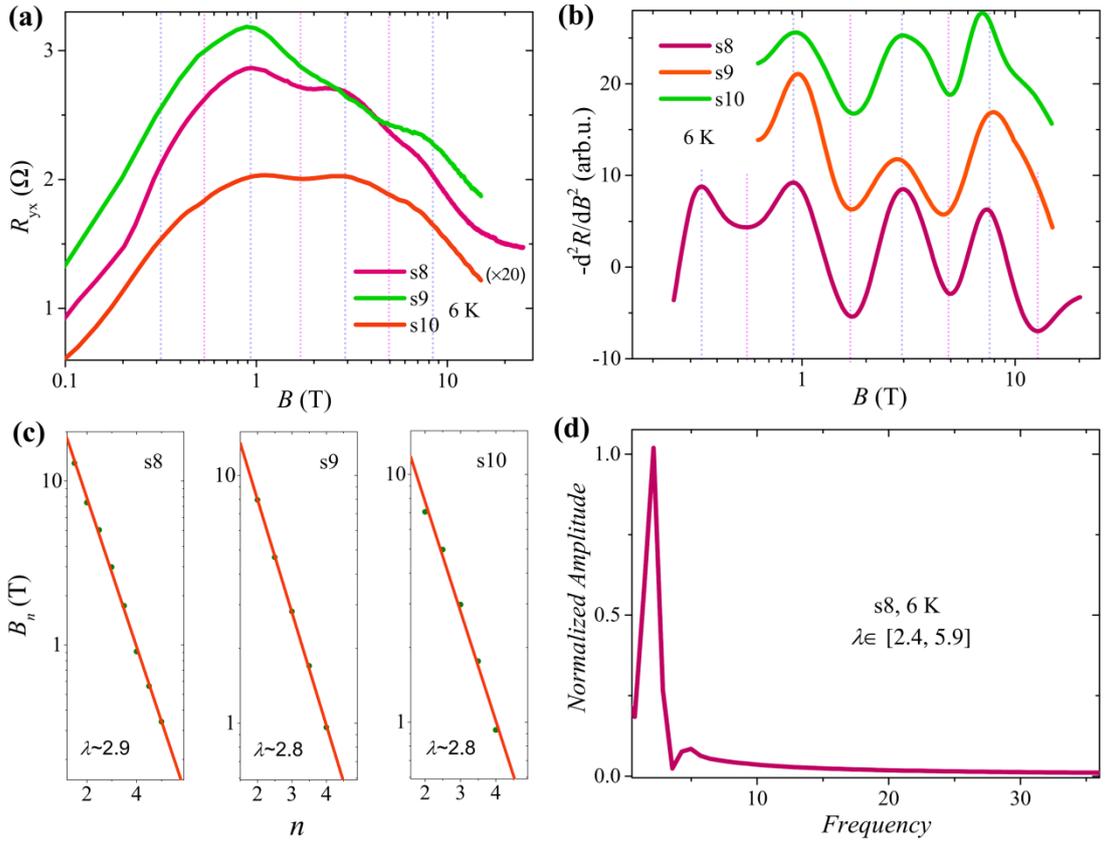

FIG. 4 (color online). Signals of log$B$-periodic oscillations in the Hall traces of HfTe$_5$. (a) Hall traces of HfTe$_5$ crystals at 6 K versus the magnetic field. (b) The second derivative results of the curves in (a). Data curves are shifted for clarity. (c) The log-periodicity of the oscillating Hall resistance. (d) FFT result of the oscillations in s8. Based on the FWHM of the FFT frequency peak, the scale factor $\lambda$ shows an approximate range of [2.4, 5.9] consistent with the results of the MR behavior.

Supplemental Material for

# Log-periodic quantum magneto-oscillations and discrete scale invariance in topological material HfTe$_5$

Huichao Wang,[1,2] Yanzhao Liu,[1] Yongjie Liu,[3] Chuanying Xi,[4] Junfeng Wang,[3] Jun Liu,[5] Yong Wang,[5] Liang Li,[3] Shu Ping Lau,[2] Mingliang Tian,[4] Jiaqiang Yan,[6] David Mandrus,[6,7] Ji-Yan Dai,[2,*] Haiwen Liu,[8,‡] X. C. Xie,[1,9,10] and Jian Wang[1,9,10,†]

[1]*International Center for Quantum Materials, School of Physics, Peking University, Beijing 100871, China*
[2]*Department of Applied Physics, The Hong Kong Polytechnic University, Kowloon, Hong Kong, China*
[3]*Wuhan National High Magnetic Field Center, Huazhong University of Science and Technology, Wuhan 430074, China*
[4]*High Magnetic Field Laboratory, Chinese Academy of Sciences, Hefei 230031, Anhui, China*
[5]*Center of Electron Microscopy, State Key Laboratory of Silicon Materials, School of Materials Science and Engineering, Zhejiang University, Hangzhou, 310027, China*
[6]*Materials Science and Technology Division, Oak Ridge National Laboratory, Oak Ridge, Tennessee 37831, USA*
[7]*Department of Materials Science and Engineering, University of Tennessee, Knoxville, Tennessee 37996, USA*
[8]*Center for Advanced Quantum Studies, Department of Physics, Beijing Normal University, Beijing, 100875, China*
[9]*Collaborative Innovation Center of Quantum Matter, Beijing 100871, China*
[10]*CAS Center for Excellence in Topological Quantum Computation, University of Chinese Academy of Sciences, Beijing 100190, China*

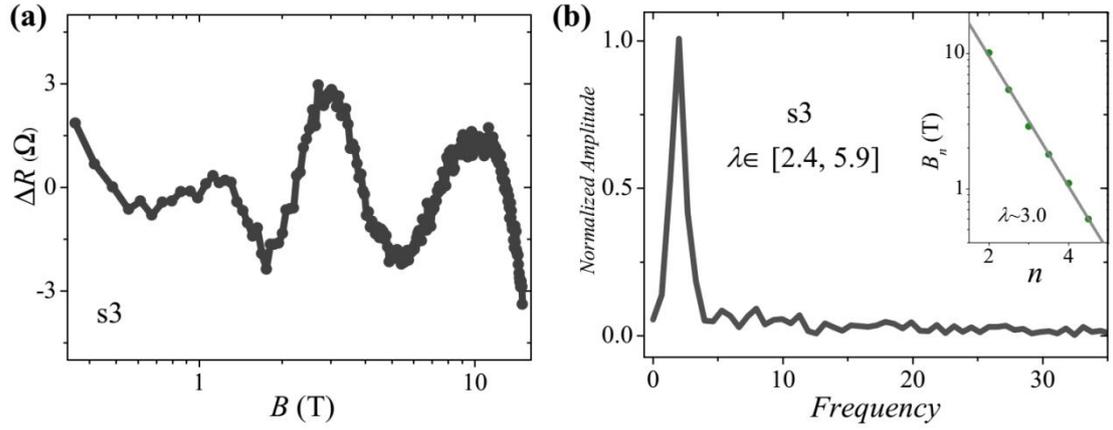

FIG. S1 (a), The magneto-resistance (MR) oscillations in s3 by subtracting a background. (b), The FFT result of (a) confirms the oscillations periodic in log$B$. Inset: the index plot also shows log-periodicity of the oscillations. From the FWHM of the FFT frequency peak, a range of [2.4, 5.9] is obtained for the scale factor $\lambda$. Thus, the results by subtracting background are consistent with results of the second derivative in Fig.1 in the main text and we demonstrate that the oscillations are intrinsic in HfTe$_5$.

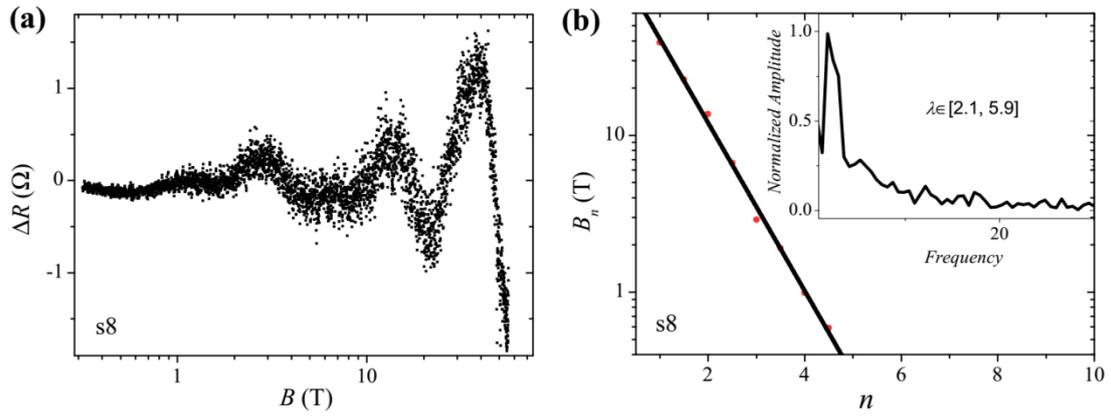

FIG. S2 (a), Extracted MR oscillations in s8 at 4.2 K. (b), The log-periodicity of the MR oscillations. Inset: FFT result of the oscillations in (b). An approximate range for the scale factor is [2.1, 5.9], which is determined by the FWHM of the FFT frequency peak.

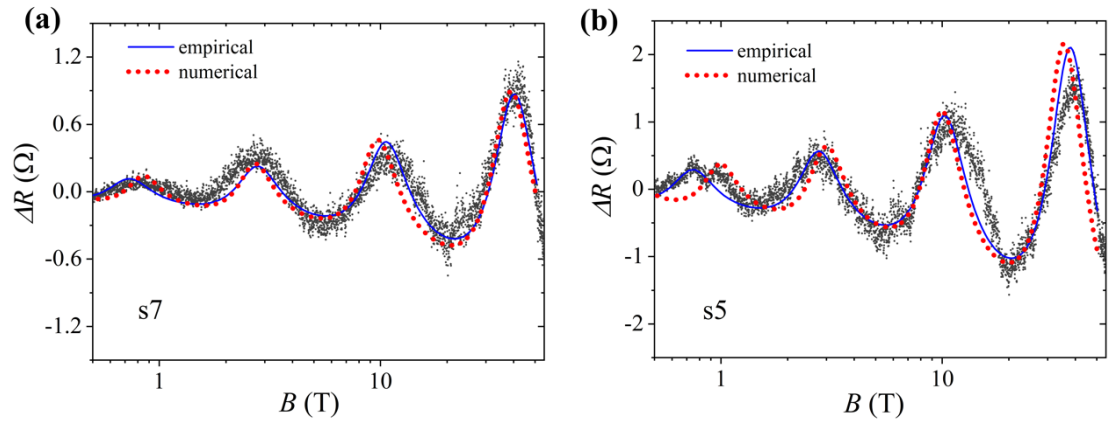

FIG. S3. The fitting of experimental data based on the microscopic numerical calculation and the empirical formula for s7 at 4.2 K (a) and s5 at 3.0 K (b).

Sample Information

The samples s1-s4 were only measured in PPMS for magneto-resistance (MR) behavior. The MR of s5 was measured in both the PPMS and the pulsed high magnetic field up to 58 T. The MR of s6 and s7 were only measured in the pulsed high magnetic field. The sample s8 was measured in the pulsed magnetic field up to 58 T to explore the MR oscillations while the Hall trace of s8 was only measured in the static high magnetic field up to 25 T in Hefei for better signal. The s9 and s10 was only measured in PPMS for detecting Hall traces. It is noted that the results for the sample from different measurement systems are consistent and the log$B$-periodic oscillations appear with similar scale factor range in these different samples.

Theoretical Details

Based on the t-matrix approximation, we derive the longitudinal conductivity beyond the quantum limit under large magnetic fields (the details are given in ref. 14 and the theoretical preprint [1]):

$$\sigma_{xx}(\varepsilon_F) = \frac{4e^2}{h} l_B^2 \left( n_s + n_c \frac{t^2}{8\pi \cdot \hbar v_F l_*^{-1} \cdot \Gamma(B)} \sum_n \frac{\Gamma(B)^2}{(\varepsilon_F - \varepsilon_n(B))^2 + \Gamma(B)^2} \right) \quad (SE1).$$

Here $n_s$ is the density of short-range scatterers, $n_c$ is the density of Coulomb scatterers, $l_*$ is the effective length along the magnetic field, $t$ is the coupling strength between the bound states with the continuum of the lowest Landau level, $\varepsilon_F$ is the Fermi energy, $\varepsilon_n(B)$ is the energy for the $n$-th bound state and $\Gamma(B) = \eta^* \sqrt{B}$ is the width mainly determined by the broadening effect of the lowest Landau level with $\eta^*$ depending on the microscopic scattering process and temperature [2]. The above microscopic formula can be further simplified into an empirical form which are more suitable for fitting the experimental data:

$$\sigma_{xx} = \frac{4e^2}{h} l_B^2 \left( n_s + n_c \frac{t^2}{8\pi \cdot \hbar v_F l_*^{-1} \cdot \Gamma(B)} \frac{\eta^2}{sin^2\left(\frac{s_0}{2} ln\left(\frac{B}{B_0}\right)\right) + \eta^2} \right) \quad (SE2),$$

$$\sigma_{xy} = \frac{4e^2}{h} l_B^2 \left( N + n_c \frac{t^2}{8\pi \cdot \hbar v_F l_*^{-1}} \sum_n \frac{\varepsilon_F - \varepsilon_n(B)}{(\varepsilon_F - \varepsilon_n(B))^2 + \Gamma(B)^2} \right) \quad (SE3).$$

Here, $N$ denotes the total carrier density, $s_0$, $B_0$ and $\eta$ are fitting parameters. In equation (SE2), the first term denotes the Anderson impurity scattering of the mobile carriers, which leads to linear-B dependent MR previously obtained by A. A. Abrikosov [3]; and the second term denotes the resonant scattering between the mobile carriers and the quasi-bound states, which gives rise to log-periodic correction to the MR. The correction to $\rho_{xx}$ is of the order $n_C/n_S$, while that to $\rho_{yx}$ is of the order $n_C/N$.

The microscopic formula eq. (SE1) and the empirical formula eq. (SE2) can be used for fitting the experimental results. Figure S3 shows the fitting curves obtained with three fitting parameters $s_0$, $B_0$ and $\eta^*$ (for SE1)/ $\eta$ (for SE2). Here, $s_0$ and $B_0$ determine the energy for the $n$-th bound state $\varepsilon_n(B)$ in the microscopic formula eq. (SE1). At different temperature, the thermal broadening effect changes the fitting parameters $\eta^*$ or $\eta$. The fitting parameters are $s_0$=4.8 (4.7) and $B_0$=0.20 T (0.19 T) for sample s5 (s7). These parameters change slightly in different samples, which is consistent with the expectation that the DSI feature is mainly determined by fine structure constant in the material HfTe$_5$.

[*]jiyan.dai@polyu.edu.hk;
[‡]haiwen.liu@bnu.edu.cn;
[†]jianwangphysics@pku.edu.cn